\documentclass[a4paper,11pt]{article}
\usepackage{jinstpub}
\usepackage{graphicx}

\ifdefined\pdfcompresslevel
\fi
\ifdefined\pdfobjcompresslevel
\fi

\graphicspath{{figures/}}
\DeclareGraphicsExtensions{.pdf,.jpg,.jpeg,.png}

\newif\ifOverleafDraft
\OverleafDraftfalse
\ifOverleafDraft
  \setkeys{Gin}{draft}
\fi

\title{\boldmath Fabrication and characterization of lithium-diffused inverted coaxial point-contact HPGe detectors}

\author[a]{Anupama Karki,}
\author[a]{Kunming Dong,}
\author[a,1]{and Dongming Mei\note{Corresponding author}}

\affiliation[a]{Department of Physics, University of South Dakota,\\
414 E. Clark St., Vermillion, SD 57069, U.S.A.}

\emailAdd{dongming.mei@usd.edu}

\abstract{
High-purity germanium (HPGe) detectors with low capacitance, low electronic noise, and stable high-voltage operation are important for high-resolution gamma-ray spectroscopy and rare-event searches. We report the fabrication and characterization of two compact p-type inverted coaxial point-contact (ICPC) HPGe detectors, AK01 and AK02, produced from crystals grown at the University of South Dakota. Both devices use lithium-diffused $n^{+}$ outer contacts together with amorphous-Ge/Al point-contact electrodes. The principal advance is the implementation of this hybrid contact process on two independently fabricated USD-grown ICPC prototypes and the demonstration of stable electrical and spectroscopic operation. Both detectors exhibited picoampere-level leakage currents, operational depletion-voltage ranges of approximately 200--250~V (AK01) and 240--260~V (AK02), and estimated effective plateau capacitances of approximately 0.84 and 0.87~pF, respectively. Measurements with a $^{137}$Cs source yielded archived full-energy-peak FWHM values of 1.71 and 1.62~keV near 662~keV for AK01 and AK02, while AK02 yielded 1.04~keV FWHM at 59.5~keV with $^{241}$Am. Electric-field calculations reproduce the expected ICPC field configuration for the measured detector geometry. A Geant4 model of AK02 is used as a qualitative response study; its 0.8-mm Li-diffused inactive layer is an illustrative model assumption rather than a measured thickness, and its Gaussian broadening is constrained by the experimental resolution. The work therefore establishes the fabrication and operation of compact lithium-diffused ICPC prototypes while identifying quantitative inactive-layer characterization as an important next step.
}

\keywords{Gamma detectors (scintillators, CZT, HPGe, HgI etc); Solid state detectors; Detector modelling and simulations; Instrumentation for gamma-ray spectroscopy studies}

\begin{document}
\maketitle
\raggedbottom
\section{Introduction}

High-purity germanium (HPGe) detectors are among the most widely used radiation detectors for gamma-ray spectroscopy because of their excellent energy resolution, high detection efficiency, and low intrinsic electronic noise \cite{knoll2010}. These characteristics make HPGe detectors a leading technology for applications in nuclear physics, homeland security, environmental monitoring, and rare-event searches. In particular, experiments searching for neutrinoless double-beta decay, such as the GERDA \cite{gerda2020}, Majorana Demonstrator \cite{majorana2014}, and LEGEND collaborations \cite{legend2017}, rely on HPGe detectors to achieve the stringent energy resolution and low-background requirements necessary for detecting extremely rare interactions.

Point-contact HPGe detector geometries have attracted considerable attention because they provide exceptionally low detector capacitance, resulting in reduced electronic noise and superior spectroscopic performance \cite{luke1989,barbeau2007}. The inverted coaxial point-contact (ICPC) geometry combines the
low-capacitance characteristics of point-contact detectors with a
geometry that can accommodate relatively large detector volumes~\cite{salathe2017,gerda2021}.
This configuration is particularly attractive for low-background
experiments because it preserves good energy resolution and
pulse-shape discrimination while increasing the detector mass
~\cite{yang2023,gerda2021,comellato2021}. ICPC detectors have been developed and operated
in GERDA and are relevant to the detector technology employed in
next-generation germanium-based rare-event searches~\cite{legend2017,gerda2021}.
The performance of HPGe detectors is strongly influenced by detector geometry, impurity concentration, electrical contacts, and dead-layer formation~\cite{knoll2010}. Electrical contacts are particularly critical because they determine charge collection efficiency, leakage current, and long-term detector stability~\cite{bhattarai2020,amman2018,luke1992}. In p-type HPGe detectors, lithium-diffused $n^{+}$ contacts are commonly
used to form the outer electrode~\cite{majorana2013,fuller1953}. The lithium-diffused region can contain
an inactive or partially active layer near the detector surface, which can
influence charge collection and reduce the effective active detector
volume~\cite{dai2023,majorana2013,ma2017}. The properties of this region are therefore
important for understanding detector response, particularly for
low-energy radiation~\cite{dai2023,ma2017}.

Lithium diffusion in germanium has been studied extensively for several decades, and diffusion coefficients have been reported under various thermal processing conditions~\cite{fuller1953,pell1957,carter1960}. Nevertheless, the resulting diffusion depth and dead-layer thickness depend strongly on fabrication parameters such as diffusion temperature, diffusion duration, and cooling history. Accurate control of the lithium diffusion process is therefore essential for optimizing detector performance and maximizing active detector volume.

Recent efforts at the University of South Dakota have focused on the development of HPGe detector technologies using locally grown crystals~\cite{wang2015}. As part of this effort, fabrication and characterization studies are being performed to establish reliable detector processing procedures and evaluate the impact of contact fabrication methods on detector performance. In particular, understanding the relationship between lithium diffusion, detector depletion characteristics, leakage current, and spectroscopic response is important for future development of advanced germanium detector geometries.

In this work, two lithium-diffused inverted coaxial point-contact HPGe
detectors, designated AK01 and AK02, were fabricated and characterized at
the University of South Dakota. The emphasis is not on establishing lithium diffusion as a new contact concept, but on demonstrating the hybrid-contact fabrication route on two independently fabricated USD-grown ICPC crystals. Relative to our
previous a-Ge-contact ICPC work~\cite{panamaldeniya2026}, the present study provides four
specific advances: (i) implementation of a lithium-diffused $n^{+}$ outer
contact together with an a-Ge/Al point-contact electrode on compact ICPC
geometries; (ii) successful operation of two independently fabricated
prototypes with picoampere-level leakage current and sub-pF effective
capacitance; (iii) combined electrical, spectroscopic, and electrostatic
characterization of the fabricated devices; and (iv) an explicitly
limited Geant4 response study that identifies the Li-diffused inactive
layer as a key quantity requiring future calibrated efficiency measurements.
The principal result is therefore the demonstrated fabrication and operation
of the two hybrid-contact ICPC prototypes rather than lithium diffusion
itself.

\section{Detector fabrication}

\subsection{Detector geometry and contact structure}

The AK01 and AK02 detectors were fabricated from p-type HPGe crystals grown
at the University of South Dakota using the Czochralski method. A Hall-effect
measurement on one side of the crystal indicated a net impurity concentration
of approximately $5\times10^{10}\ \mathrm{cm^{-3}}$, while the opposite side
had a lower impurity concentration. These Hall measurements indicate impurity
variation within the crystal and are not used here to assign a unique,
detector-specific net impurity concentration to either AK01 or AK02.

\begin{figure}[htbp]
\centering
\includegraphics[width=0.65\textwidth]{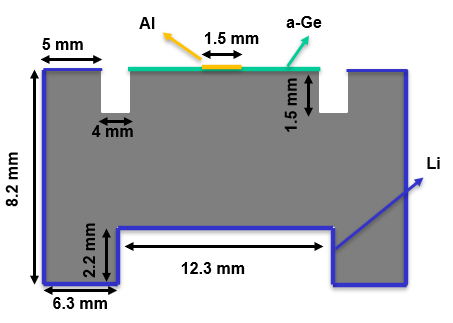}
\caption{Schematic cross-sectional view of the AK01 lithium-diffused ICPC HPGe detector.}

\label{fig:icpc}
\end{figure}

Both detectors employed an inverted coaxial point-contact geometry with the
same general contact configuration. Each detector consisted of a cylindrical
crystal with a coaxial bore machined from the bottom surface. A circumferential
groove surrounding the point-contact region provided electrical isolation
between the point-contact electrode and the outer contact. A lithium-diffused $n^{+}$ contact was formed on the outer cylindrical
surface and the bore region, while an amorphous-Ge blocking contact and
an Al readout electrode were deposited on the point-contact region.
Lithium diffusion in germanium has long been used to form highly doped
$n^{+}$ regions~\cite{fuller1953}, whereas amorphous-Ge contacts are widely
employed in HPGe detectors as carrier-injection blocking and
surface-passivating contacts~\cite{amman2020,luke1992}. Although the two detectors share the same overall design, they differ in detector dimensions, bore geometry, groove dimensions, and mass. The geometrical and fabrication parameters of AK01 and AK02 are summarized in Table~\ref{tab:geometry}. Figure~\ref{fig:icpc} shows a schematic cross-sectional view of the lithium-diffused ICPC detector geometry used in this work.
 
 \begin{table}[htbp]
\centering
\caption{Geometrical and fabrication parameters of the lithium-diffused ICPC HPGe detectors.}
\begin{tabular}{lcc}
\hline
Parameter & AK01 & AK02 \\
\hline
Crystal material & p-type HPGe & p-type HPGe \\
Detector body diameter & 24.9 mm & 25.3 mm \\
Detector body height & 8.2 mm & 9.4 mm \\
Coaxial bore diameter & 12.3 mm & 15.6 mm \\
Coaxial bore depth & 2.2 mm & 2.1 mm \\
Circumferential groove width & 4.0 mm & 4.9 mm \\
Al point-contact electrode diameter & 1.5 mm & 1.5 mm \\
Detector mass & 18 g & 20 g \\
\hline
\end{tabular}
\label{tab:geometry}
\end{table}

\subsection{Crystal machining and surface preparation}

\begin{figure}[htbp]
\centering\includegraphics[width=0.85\textwidth]{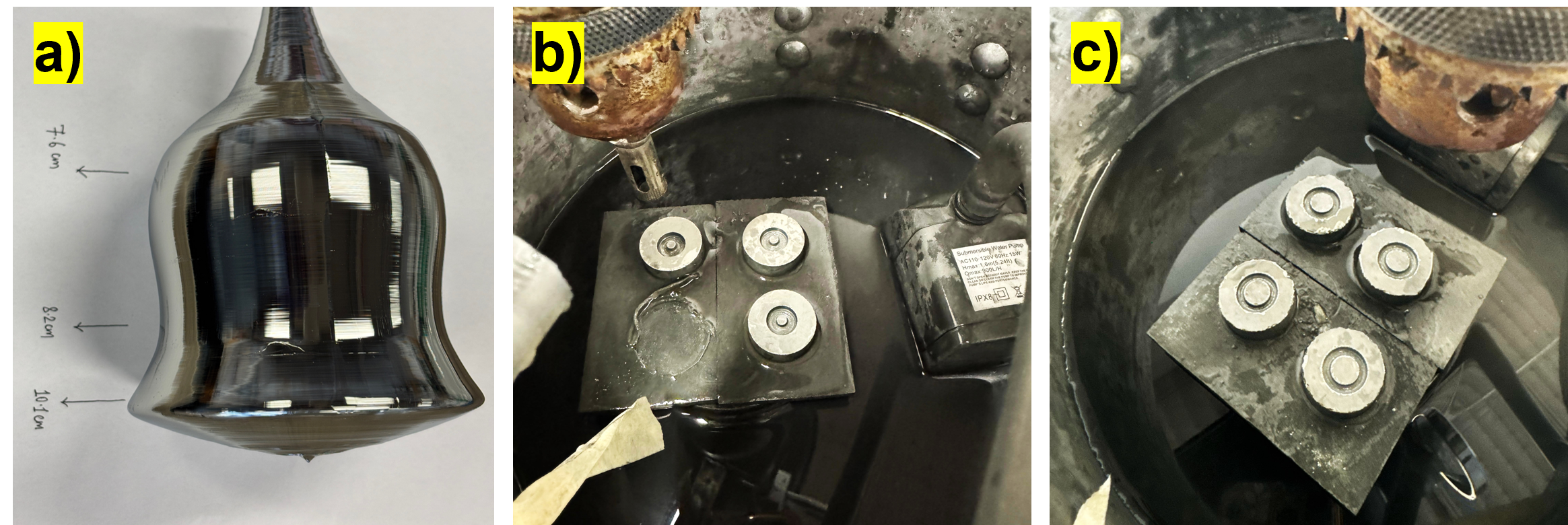}
\caption{Several stages of the ICPC detector mechanical fabrication process: (a) germanium crystal ingot, (b) coaxial bore machining, and (c) groove and point-contact formation.}

\label{fig:Cutting}
\end{figure}

The detector fabrication procedure followed methods similar to those reported in~\cite{panamaldeniya2026}. The inverted coaxial point-contact geometry was
machined from cylindrical HPGe samples using a combined lathe, mill, and
drill system. The machining steps included shaping the cylindrical detector
body, forming the coaxial bore and circumferential groove, and defining the
point-contact region.  Figure~\ref{fig:Cutting} shows several stages of the ICPC mechanical fabrication process.

\begin{figure}[htbp]
\centering
\includegraphics[width=0.85\textwidth]{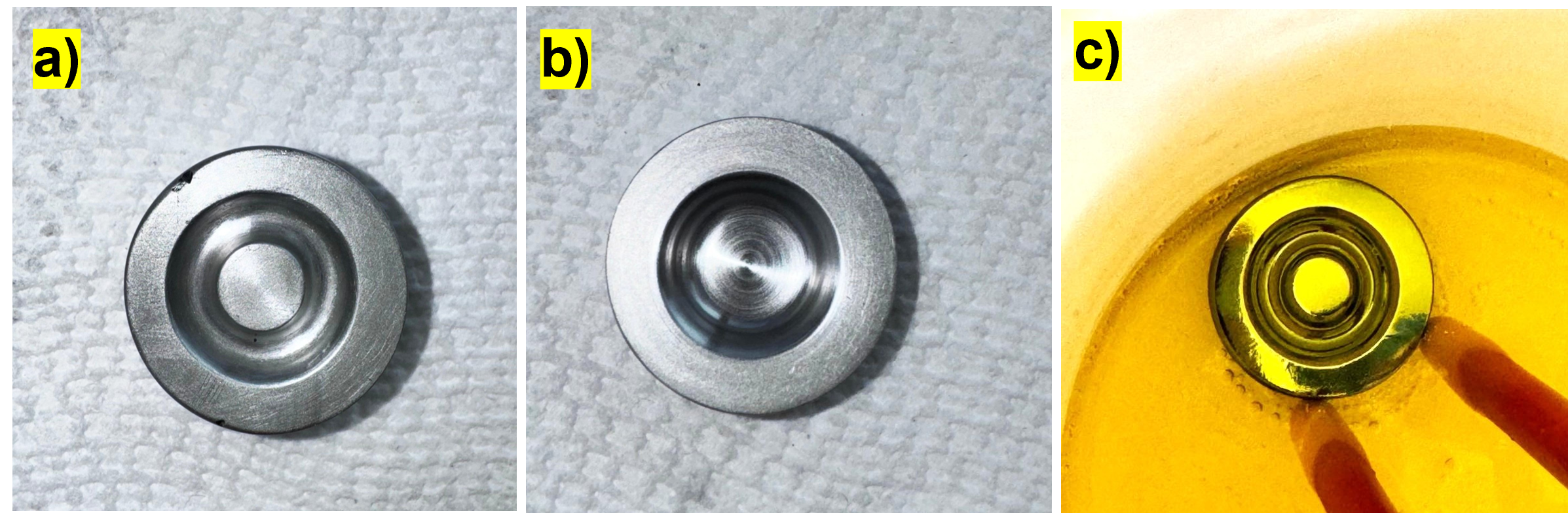}

\caption{Surface preparation stages of the ICPC HPGe detector: (a) polished groove and point-contact surface, (b) polished bore-hole surface, and (c) chemical etching.}

\label{fig:polishing}
\end{figure}

Following mechanical fabrication, the detector surfaces were prepared through a sequence of lapping, polishing, and chemical etching steps. The top and bottom detector faces were lapped on a glass plate using abrasive mixtures prepared from 17.5~$\mu$m SiC powder and deionized (DI) water for coarse lapping, followed by 9.5~$\mu$m Al$_2$O$_3$ powder and DI water for fine lapping, while the cylindrical sidewall and inverted coaxial bore surfaces were polished using successive SiC abrasive papers with grit sizes of 120, 240, 400, 800, 1000, 1250, 1500 and 2500. After polishing, the detectors were etched in an HF:HNO$_3$ (1:4) solution for approximately 6~min with continuous agitation to remove residual subsurface damage and surface contaminants. Immediately following etching, the detectors were thoroughly rinsed in deionized (DI) water and dried under nitrogen flow.
Figure~\ref{fig:polishing} shows the detector surfaces after polishing and chemical etching.

\subsection{Lithium deposition and diffusion}

Prior to lithium deposition, the detector surfaces were chemically etched to remove residual machining damage and surface contamination introduced during cutting, lapping, and polishing~\cite{wei2022,amman2020}. Following etching, the detectors were dried with high-purity nitrogen and transferred into an argon glovebox through an antechamber pump-and-backfill cycle. The low-oxygen environment minimizes oxidation of freshly etched germanium surfaces and helps maintain surface cleanliness during subsequent lithium deposition and diffusion processes~\cite{wei2022,amman2020}. Inside the glovebox, the detectors were mounted on a cleaned graphite holder and positioned on the heater platform for lithium deposition and thermal diffusion.

Following surface preparation and chemical etching, lithium was deposited on the designated detector surfaces using a lithium suspension paint prepared according to the procedure described in Ref.~\cite{dong2026}. Before application, the suspension was thoroughly mixed to ensure a uniform lithium distribution. A thin and uniform lithium layer was applied using a fine brush to the outer detector surfaces and the inner bore region designated for the formation of the $n^{+}$ contact.

After lithium deposition, the detectors were placed on a graphite holder inside an argon glovebox and subjected to the same nominal thermal diffusion recipe. During diffusion, the detector surface temperature was maintained at approximately 280~$^\circ$C for about 30~min under an argon atmosphere to allow lithium diffusion into the germanium crystal. These temperature and time values are process conditions; the present study does not use them to infer a unique lithium-junction depth because the final profile can also depend on the deposited Li distribution, thermal gradients, solubility/precipitation behavior, and cooling history~\cite{fuller1953,pell1957,carter1960}. No independent Li concentration-depth profile or inactive-layer thickness was measured for AK01 or AK02.

\begin{figure}[htbp]
\centering
\includegraphics[width=0.85\textwidth]{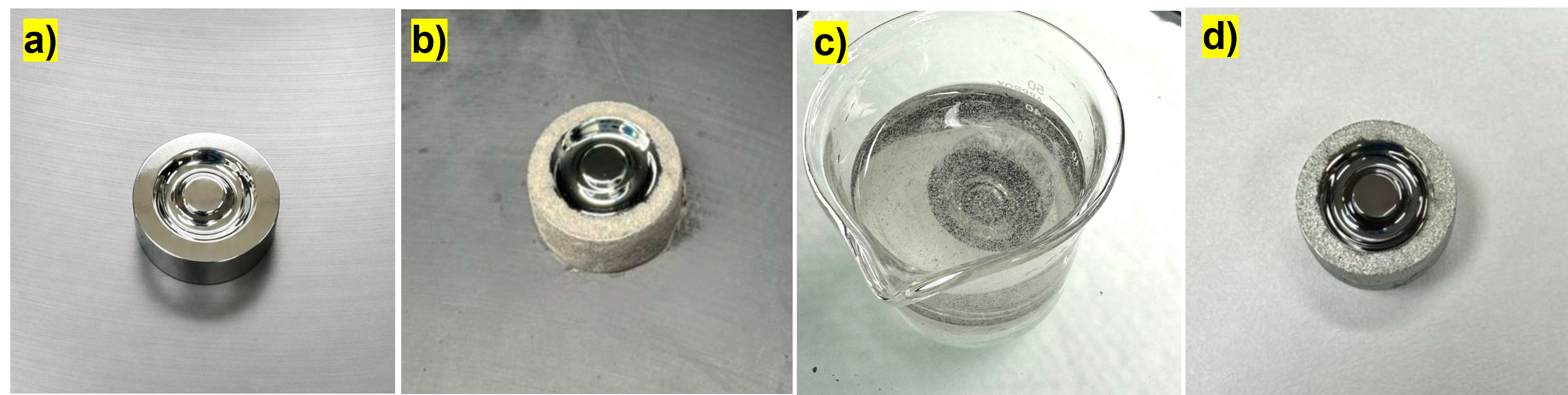}
 \caption{Lithium contact fabrication process for the ICPC detector: (a) detector prior to lithium deposition, (b) lithium paint applied to the detector surface, (c) detector immersed in methanol immediately after diffusion with visible surface bubbling, and (d) lithium-diffused detector after cleaning.}

    \label{fig:Lidiffusion}
\end{figure}

Following diffusion, the detectors were rapidly cooled on an aluminum block to minimize additional lithium redistribution during cooling. The cooling step is therefore part of the fabrication recipe, but the present work does not assign a quantitative diffusion depth from the cooling transient. After cooling to room temperature, the detectors were removed from the glovebox and immersed in methanol to remove residual lithium from the diffused surfaces. Vigorous bubbling was observed during this process as the residual lithium reacted with the methanol. The detector surfaces were gently brushed to assist lithium removal. After the bubbling ceased, the detectors were transferred to isopropyl alcohol (IPA), rinsed in deionized (DI) water, and dried with high-purity nitrogen. A brief HNO$_3$:HF (4:1) etch ($\sim$30 s) was then performed to remove the reacted surface layer and refresh the germanium surface prior to contact deposition~\cite{wei2022,amman2020}. The detectors were subsequently rinsed in DI water, dried with nitrogen, and prepared for a-Ge and Al contact deposition. Figure~\ref{fig:Lidiffusion} illustrates several stages of the lithium deposition and diffusion process.

\subsection{a-Ge and Al contact deposition}

Following lithium diffusion and surface preparation, amorphous germanium (a-Ge) and aluminum (Al) contacts were deposited using an AJA magnetron sputtering system. Amorphous-Ge contacts are widely used in HPGe detectors because they provide effective charge-blocking behavior and stable operation at cryogenic temperatures~\cite{bhattarai2020}. Prior to deposition, the sputtering chamber was evacuated to a base pressure of approximately $3\times10^{-7}$ Torr to minimize residual contamination. The a-Ge target was pre-sputtered before deposition to clean the target surface and stabilize the plasma~\cite{amman2020}.

The a-Ge layer was deposited using a 7\% H$_2$/Ar process gas with a total flow of 20 sccm at a chamber pressure of 3 mTorr and an RF power of 200 W. During deposition, the detector stage was rotated at 2 rpm to improve film uniformity. A deposition time of 15 min produced an a-Ge film thickness of approximately 370 nm. The a-Ge layer was deposited only on the point-contact region, where it serves as a charge-blocking contact. The circumferential groove was intentionally left uncoated to maintain electrical isolation between the point-contact electrode and the lithium-diffused outer contact.

\begin{figure}[htbp]
\centering
\includegraphics[width=0.85\textwidth]{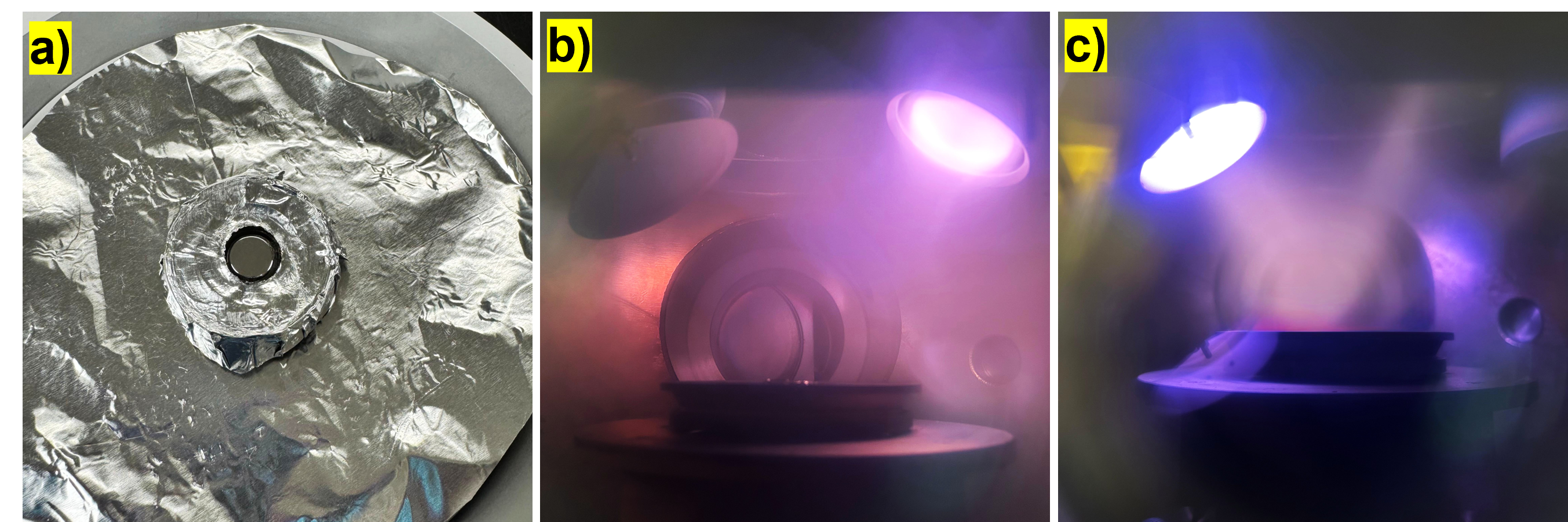}
\caption{a-Ge and Al contact deposition process using the AJA sputtering system: (a) detector mounted and masked prior to deposition, (b) RF magnetron sputtering of the a-Ge contact and (c) DC magnetron sputtering of the Al readout electrode.}
\label{fig:sputtering}
\end{figure}

Without breaking vacuum, an aluminum layer was subsequently deposited on top of the a-Ge contact using DC magnetron sputtering in pure argon. The deposition was performed at a chamber pressure of 3 mTorr for approximately 6 min. A DC power of 400 W was used for AK01, while AK02 employed a DC power of 300 W. The two values correspond to separate fabrication runs; no controlled sputtering-power study was performed, so this run-to-run process difference is not used to explain differences in detector performance. The resulting Al layer provided a low-resistance electrical interface for signal readout through the point-contact electrode. Figure~\ref{fig:sputtering} shows representative stages of the a-Ge and Al contact deposition process.

\subsection{Point-contact formation}

\begin{figure}[htbp]
\centering
\includegraphics[width=0.85\textwidth]{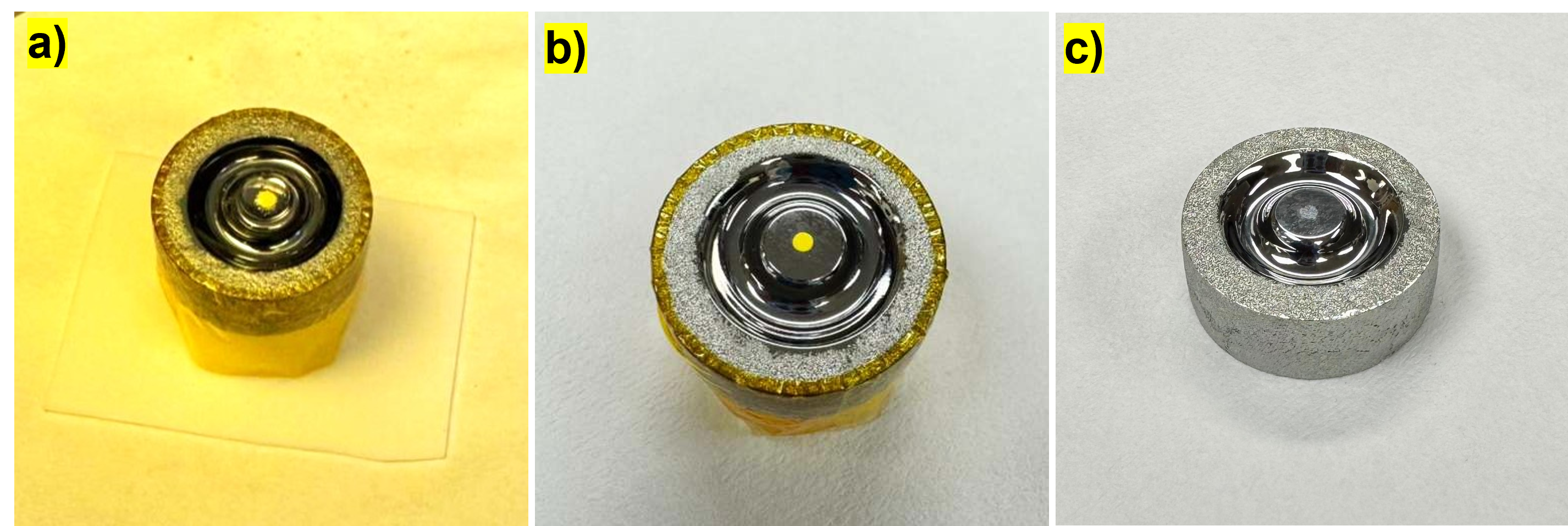}
\caption{Point-contact formation process: (a) HF etching of the point-contact region, (b) point-contact region after etching, and (c) completed point-contact electrode after cleaning and drying.}
\label{fig:contact}
\end{figure}

The point-contact electrode was defined by selectively removing the aluminum surrounding the central contact region using a localized 100:1 HF dip (approximately 0.49\% HF) applied through a Kapton masking template. The etching process was performed for approximately 3 min, leaving a small central Al/a-Ge electrode with a measured diameter of approximately 1.5 mm, determined using a caliper. This procedure removed unwanted Al from the unmasked region while preserving the desired point-contact electrode. The resulting contact was electrically isolated from the lithium-diffused outer contact by the surrounding groove region. Following etching, the detector was rinsed, cleaned, and dried prior to electrical characterization.
 Figure~\ref{fig:contact} shows representative stages of the point-contact formation process.

\section{Experimental setup}
\label{sec:experimental_setup}

\begin{figure}[htbp]
\centering
\includegraphics[width=0.65\textwidth]{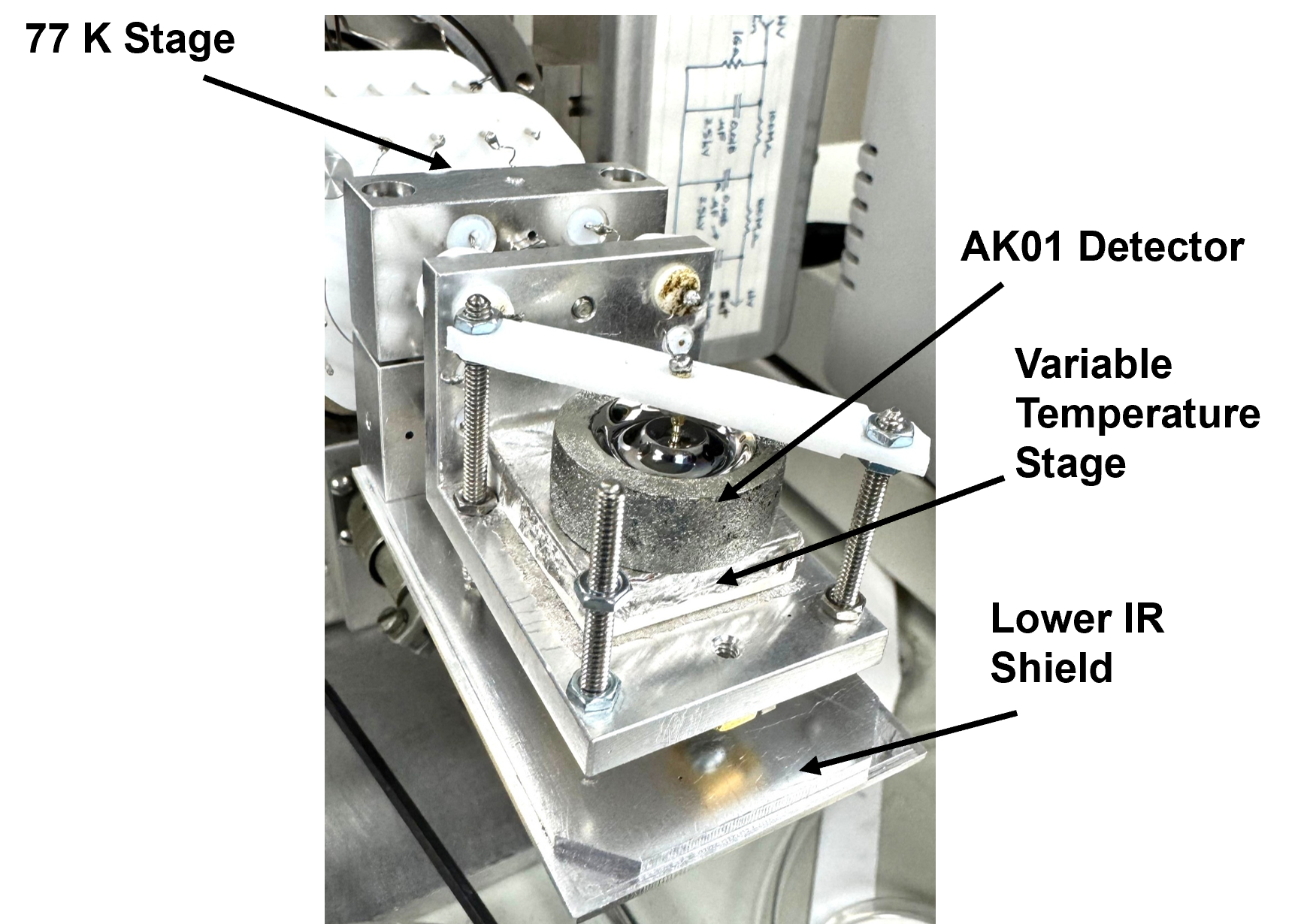}
\caption{Photograph of the fabricated ICPC detector mounted inside the cryostat during electrical and spectroscopic characterization.}
\label{fig:cryostat}
\end{figure}

Detector characterization was performed using the cryogenic test system and electronics chain developed at the University of South Dakota. The measurement configuration consisted of a high-voltage bias supply, charge-sensitive preamplifier, shaping amplifier, and multichannel analyzer (MCA) for spectroscopic measurements. Leakage-current measurements were performed using a Keithley 6482 dual-channel picoammeter, while the effective detector capacitance was estimated using a
pulser-based charge-injection method. A detailed description of the cryostat and electronics configuration is provided in Ref.~\cite{panamaldeniya2026}. All measurements were carried out with the detector operated at liquid-nitrogen temperature. A positive high-voltage bias was applied to the lithium-diffused outer contact, while the point-contact electrode was maintained at ground potential and used for signal readout. For spectroscopic measurements, a shaping time of
$0.5\,\mu\mathrm{s}$ was used for the $^{137}\mathrm{Cs}$
measurements, while $1\,\mu\mathrm{s}$ was used for the
$^{241}\mathrm{Am}$ measurement. These shaping times were selected to provide stable signal processing and good spectroscopic performance throughout detector characterization. Figure~\ref{fig:cryostat} shows the fabricated detector mounted inside the cryostat during characterization.

\subsection{Capacitance measurement method}
\label{sec:capacitance_method}

The capacitance response of each detector was evaluated using a
pulser-based charge-injection technique. A fixed-amplitude test-pulser
signal was introduced through the preamplifier test input, and the
corresponding pulse amplitude was monitored using an oscilloscope.
The MCA energy scale was calibrated using the 661.7~keV photopeak from
$^{137}$Cs and the 59.5~keV photopeak from $^{241}$Am. The centroid
position of the pulser peak in the calibrated MCA spectrum was then used
to determine its equivalent energy, $E_{\mathrm{p}}$.

The injected charge corresponding to the equivalent pulser energy was
calculated as

\begin{equation}
Q = Ne = \frac{E_{\mathrm{p}}}{w}e,
\label{eq:injected_charge}
\end{equation}

where $N$ is the equivalent number of electron--hole pairs, $e$ is the
elementary charge, and $w=2.96$~eV is the mean energy required to produce
one electron--hole pair in Ge at 77~K~\cite{knoll2010}.

The measured pulser amplitude was corrected using the attenuation factor
of the measurement circuit,

\begin{equation}
\alpha
=
\frac{10}{10+44}
=
0.1852,
\label{eq:attenuation_factor}
\end{equation}

such that

\begin{equation}
V_{\mathrm{corr}}
=
\alpha V_{\mathrm{pulse}},
\label{eq:voltage_correction}
\end{equation}

where $V_{\mathrm{pulse}}$ is the measured pulser amplitude and
$V_{\mathrm{corr}}$ is the corresponding corrected voltage used in the
capacitance calculation.

The effective capacitance in the high-voltage plateau region was estimated
from

\begin{equation}
C_{\mathrm{dep}}
=
\frac{Q}{V_{\mathrm{corr}}}.
\label{eq:depleted_capacitance}
\end{equation}

The bias-dependent effective capacitance was then obtained by scaling the
relative capacitance response to its plateau value:

\begin{equation}
C_{\mathrm{eff}}(V)
=
C_{\mathrm{dep}}
\frac{C_{\mathrm{rel}}(V)}
     {C_{\mathrm{rel,dep}}},
\label{eq:effective_capacitance}
\end{equation}

where $C_{\mathrm{rel}}(V)$ is the relative capacitance response at applied
bias $V$, and $C_{\mathrm{rel,dep}}$ is the corresponding value in the
high-voltage plateau region.

Because the absolute capacitance scale was inferred from the pulser
calibration rather than measured directly using a dedicated LCR bridge,
the reported values are treated as estimated effective capacitances.
The principal systematic limitations arise from the pulser-energy
calibration, oscilloscope voltage resolution, attenuation-factor
correction, resistor tolerances, electronic-gain stability, and parasitic
contributions from the measurement circuit. Consistent with the same
charge-injection calibration framework used in our earlier ICPC work~\cite{panamaldeniya2026},
we assign an estimated method-level relative uncertainty of approximately
6--8\% to the absolute capacitance scale. This uncertainty is larger than
the AK01--AK02 difference in the plateau values; therefore, 0.84 and
0.87~pF should be interpreted as comparable sub-pF effective capacitances
rather than as a statistically resolved difference. The full raw
pulser-amplitude/equivalent-energy pairs used in the archived calibration
are not reproduced in the present manuscript, so the equations above are
provided to document the analysis relation rather than to imply an
independent LCR-level capacitance measurement.

\section{Electrical characterization}
\subsection{Biasing configuration and leakage-current characteristics}
The fabricated ICPC detectors were operated in a liquid-nitrogen-cooled vacuum cryostat at approximately 77 K. Under the bias configuration described in Section~\ref{sec:experimental_setup}, the depletion region expands through the lightly doped p-type bulk as the applied voltage increases, reducing detector capacitance and enabling efficient charge collection \cite{knoll2010}.

At liquid-nitrogen temperature, leakage currents in HPGe detectors can arise from carrier injection at imperfect contacts, generation--recombination processes within depleted regions, and surface conduction along passivated surfaces~\cite{knoll2010,bhattarai2020,amman2018}. Consequently, leakage-current measurements provide a sensitive probe of contact quality, surface preparation, and overall detector performance.

\begin{figure}[htbp]
\centering
\includegraphics[width=0.85\textwidth]{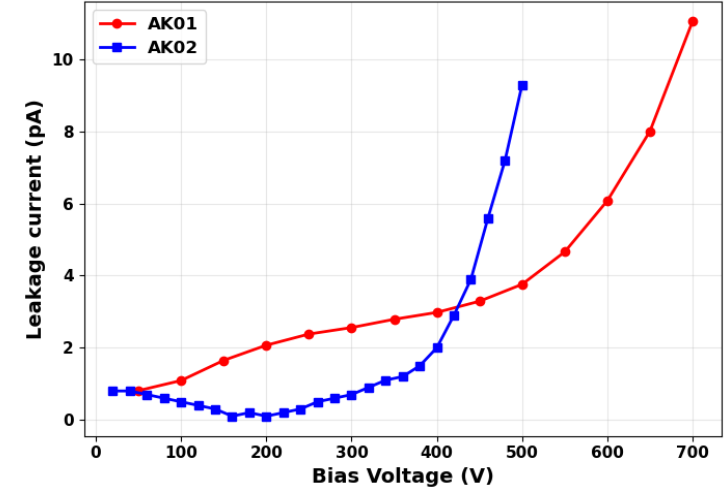}
\caption{Leakage current as a function of applied bias voltage for ICPC detectors AK01 and AK02 measured at liquid-nitrogen temperature. Both detectors exhibit picoampere-level leakage currents throughout the measured bias range.}
\label{fig:leakage}
\end{figure}

Figure~\ref{fig:leakage} shows the leakage current as a function of applied bias voltage for detectors AK01 and AK02. For AK01, the leakage current increased gradually from approximately 0.81~pA at 50~V to 11.06~pA at 700~V, which was used as the operating bias for subsequent measurements. Detector AK02 exhibited leakage currents below 1~pA at low bias voltages, followed by a more rapid increase above approximately 350--400~V, reaching 9.3~pA at 500~V, the operating bias selected for detector characterization.

For both detectors, the leakage current remained within the picoampere range throughout the measured voltage interval and at their respective operating biases. The absence of a breakdown-like discontinuity indicates stable reverse-bias operation. AK02 nevertheless shows a noticeably steeper bias dependence above approximately 350--400~V than AK01; this behavior may indicate a stronger bias-dependent contact-injection or surface-current component, although the present data do not separate those mechanisms. The overall picoampere current scale remains consistent with effective charge blocking by the lithium-diffused outer contact and a-Ge/Al point contact~\cite{bhattarai2020,majorana2013,amman2018}.

\subsection{Capacitance--voltage characteristics and
depletion-voltage determination}
\label{sec:capacitance_results}

\begin{figure}[htbp]
\centering
\includegraphics[width=0.95\textwidth]{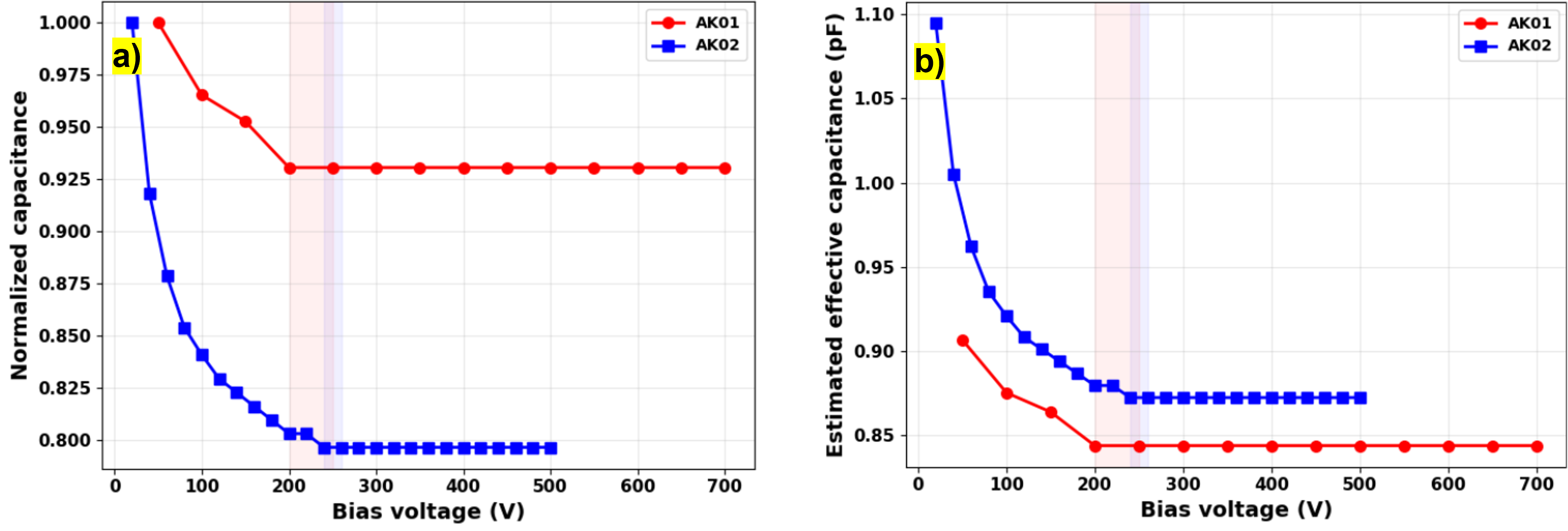}
\caption{Capacitance–voltage characteristics of the AK01 and AK02 lithium-diffused ICPC detectors. a) Normalized capacitance as a function of applied bias voltage. b) Estimated effective capacitance obtained from the pulser-based calibration. The red and blue shaded regions indicate the operational depletion-voltage ranges of 200–250 V for AK01 and 240–260 V for AK02, respectively. The capacitance becomes approximately constant above the corresponding depletion-voltage ranges.}

\label{fig:capacitance}
\end{figure}

The capacitance--voltage characteristics of AK01 and AK02 were investigated
at liquid-nitrogen temperature using the pulser-based charge-injection
method described in Section~\ref{sec:capacitance_method}.
Figure~\ref{fig:capacitance}(a) shows the capacitance response normalized
to the value measured at the lowest applied bias for each detector, while
Fig.~\ref{fig:capacitance}(b) presents the corresponding estimated effective
capacitance.

For both detectors, the capacitance decreases with increasing reverse-bias
voltage and subsequently approaches an approximately constant plateau.
At low bias, a significant fraction of the detector volume remains
undepleted, resulting in a relatively large effective capacitance. As the
applied bias increases, the depletion region expands through the detector
volume, reducing the capacitance until the high-voltage plateau is reached.
The transition to this plateau is used as an operational indicator of the onset of full
depletion~\cite{knoll2010}; it is not treated as a sharply determined physical threshold.

For AK01, the estimated effective capacitance decreases from approximately
0.91~pF at 50~V to a plateau value of approximately 0.84~pF. The transition
to the plateau occurs over approximately 200--250~V. For AK02, the estimated
effective capacitance decreases from approximately 1.09~pF at 20~V to a
plateau value of approximately 0.87~pF, with capacitance saturation occurring
near 240--260~V. These intervals are therefore reported as the operational
depletion-voltage ranges of AK01 and AK02, respectively.

For subsequent spectroscopic measurements, AK01 and AK02 were operated at
700~V and 500~V, respectively, well above their operational depletion-voltage
ranges. The comparable plateau capacitances indicate similar low-capacitance
behavior for the two prototype detectors. The normalized curves provide a
complementary representation of the depletion trend that is less sensitive
to uncertainty in the absolute pulser calibration.

\section{Gamma-ray spectroscopy}

The spectroscopic performance of the fabricated ICPC detectors was
evaluated using radioactive gamma-ray sources. A ${}^{137}\mathrm{Cs}$
source was used to characterize both AK01 and AK02, while an additional
${}^{241}\mathrm{Am}$ measurement was performed for AK02. The
${}^{137}\mathrm{Cs}$ measurements were acquired using a shaping time of
$0.5\,\mu\mathrm{s}$, whereas a shaping time of $1\,\mu\mathrm{s}$ was
used for the ${}^{241}\mathrm{Am}$ measurement. Pulser peaks were recorded
under the corresponding measurement conditions to evaluate the
contribution of the electronic readout system to the measured energy
resolution.

The reported experimental peak centroids and FWHM values are the values
retained from the archived local photopeak analyses. Formal covariance-based
fit uncertainties and the complete local background/tail model were not
retained in the manuscript record; consequently, no unsupported statistical
uncertainties are assigned here. The quoted FWHM values should therefore be
read as archived fit results, and small differences between AK01 and AK02
should not be interpreted as statistically significant without a re-fit of
the original MCA spectra.

\subsection{${}^{137}\mathrm{Cs}$ spectroscopy of AK01 and AK02}

\begin{figure}[htbp]
\centering
\includegraphics[width=0.95\textwidth]{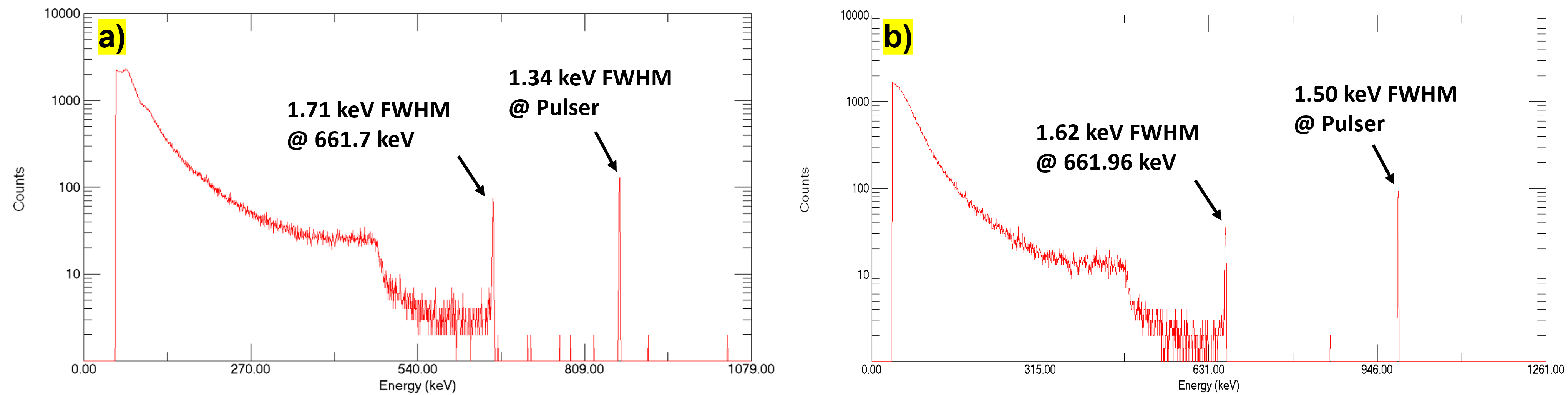}
\caption{${}^{137}\mathrm{Cs}$ energy spectra measured with
(a) AK01 and (b) AK02 using a shaping time of $0.5\,\mu\mathrm{s}$.
The measured FWHM values of the full-energy photopeaks and the
corresponding pulser peaks are indicated in the respective panels.}
\label{fig:cs137_spectra}
\end{figure}

Figure~\ref{fig:cs137_spectra} shows the ${}^{137}\mathrm{Cs}$ energy spectra measured with AK01 and AK02. Both detectors exhibit a
well-defined full-energy photopeak near $662\,\mathrm{keV}$.

For AK01, shown in figure~\ref{fig:cs137_spectra}(a), the full-energy photopeak is observed at $661.7\,\mathrm{keV}$ with a FWHM of
$1.71\,\mathrm{keV}$, corresponding to an energy resolution of
approximately $0.26\%$. The corresponding pulser peak has a FWHM of
$1.34\,\mathrm{keV}$.

For AK02, shown in figure~\ref{fig:cs137_spectra}(b), the full-energy photopeak is observed at $661.96\,\mathrm{keV}$ with a FWHM of $1.62\,\mathrm{keV}$, corresponding to an energy resolution of
approximately $0.24\%$. The corresponding pulser peak has a FWHM of
$1.50\,\mathrm{keV}$.

Both detectors therefore demonstrate comparable spectroscopic
performance at the ${}^{137}\mathrm{Cs}$ photopeak energy. The archived
fit values are 1.71 and 1.62~keV FWHM for AK01 and AK02, respectively,
but the difference is not interpreted as statistically significant because
formal peak-fit uncertainties are unavailable.

\subsection{${}^{241}\mathrm{Am}$ spectroscopy of AK02}

AK02 was additionally characterized using a ${}^{241}\mathrm{Am}$ source
to evaluate its spectroscopic response at lower gamma-ray energy. The
measurement was performed using a shaping time of $1\,\mu\mathrm{s}$.

\begin{figure}[htbp]
\centering
\includegraphics[width=0.85\textwidth]{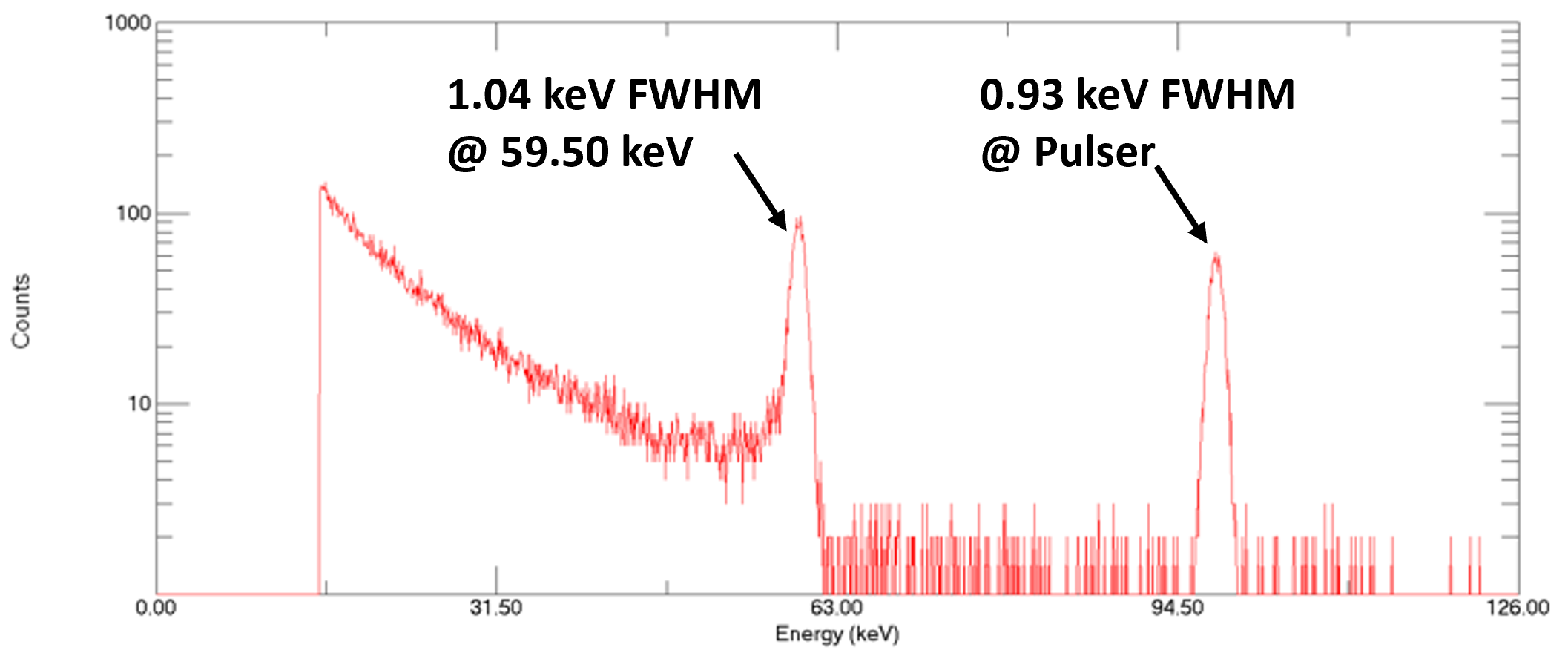}
\caption{${}^{241}\mathrm{Am}$ energy spectrum measured with AK02 using
a shaping time of $1\,\mu\mathrm{s}$. The $59.50\,\mathrm{keV}$
full-energy photopeak and the corresponding pulser peak have FWHM values
of $1.04\,\mathrm{keV}$ and $0.93\,\mathrm{keV}$, respectively.}
\label{fig:am241_ak02}
\end{figure}

Figure~\ref{fig:am241_ak02} shows the measured energy spectrum. A
well-defined full-energy photopeak is observed at $59.50\,\mathrm{keV}$
with a FWHM of $1.04\,\mathrm{keV}$, corresponding to an energy
resolution of approximately $1.75\%$. The corresponding pulser peak has
a FWHM of $0.93\,\mathrm{keV}$.

The observation of a clearly resolved $59.50\,\mathrm{keV}$ photopeak,
together with the ${}^{137}\mathrm{Cs}$ response, demonstrates the
spectroscopic operation of AK02 over both low- and higher-energy
gamma-ray measurements.

\subsection{Energy-resolution analysis}

The measured width of a gamma-ray photopeak contains contributions from
the electronic readout system as well as additional broadening associated
with the detector response. Assuming that these contributions are
independent and add in quadrature ~\cite{knoll2010}, an effective non-pulser contribution
to the measured peak width can be estimated as

\begin{equation}
\mathrm{FWHM}_{\mathrm{res}}
=
\sqrt{
\mathrm{FWHM}_{\gamma}^{2}
-
\mathrm{FWHM}_{\mathrm{pulser}}^{2}
}.
\label{eq:resolution_contribution}
\end{equation}

Applying eq.~\ref{eq:resolution_contribution} to the
${}^{137}\mathrm{Cs}$ measurements gives an effective residual
contribution of approximately $1.06\,\mathrm{keV}$ for AK01 and
$0.61\,\mathrm{keV}$ for AK02. Relative to the corresponding
full-energy photopeak energies, these values are approximately
$0.16\%$ and $0.092\%$, respectively.

For the ${}^{241}\mathrm{Am}$ measurement of AK02, the measured
photopeak and pulser widths are $1.04\,\mathrm{keV}$ and
$0.93\,\mathrm{keV}$, respectively. The same procedure gives an
effective residual contribution of approximately $0.47\,\mathrm{keV}$,
corresponding to about $0.78\%$ at $59.50\,\mathrm{keV}$.

The residual contribution obtained after quadrature subtraction may
contain several detector-dependent effects, including statistical
fluctuations in charge generation, charge trapping, incomplete charge
collection, and variations in the electric-field distribution~\cite{knoll2010,amman2018,comellato2021}.
Accordingly, the extracted quantity is treated here as an effective
non-pulser contribution to the measured energy resolution rather than
as the intrinsic statistical resolution of the detector alone. Because
the residual is obtained from the difference of two squared widths and
formal fit uncertainties are not available, the numerical residual values
are approximate and no uncertainty propagation is attempted. This caveat
is particularly important for AK02, for which the pulser width is close
to the measured gamma-ray width.

\section{Electric-field simulation}

Electric-field simulations were performed to examine the internal field
distribution of the fabricated ICPC detector geometry. The detector was
modeled using the SolidStateDetectors.jl framework~\cite{abt2021} in
cylindrical coordinates. Owing to the rotational symmetry of the ICPC
geometry, the electrostatic calculations were carried out using a
two-dimensional axisymmetric representation.

Because AK01 and AK02 have similar ICPC geometries, AK01 is presented
here as a representative example. For the electrostatic and C--V modeling,
effective uniform p-type net impurity concentrations of approximately
$1.7\times10^{10}\,\mathrm{cm^{-3}}$ for AK01 and
$1.58\times10^{10}\,\mathrm{cm^{-3}}$ for AK02 were used as modeling
parameters. These values were selected to give modeled depletion and C--V
behavior consistent with the experimentally observed operational
depletion-voltage ranges; they are not independent Hall-derived impurity
measurements for AK01 and AK02. Using
$5\times10^{10}\,\mathrm{cm^{-3}}$ with the same detector geometries would
predict substantially higher depletion voltages. The adopted values should
therefore be regarded as effective phenomenological inputs for the present
electrostatic model rather than as measured spatial impurity profiles. No
systematic impurity-concentration sensitivity scan or spatial-profile fit
was performed. The electric-field simulation presented here uses the
measured dimensions of AK01. The lithium-diffused
outer contact was treated as the positively biased electrode and the
point-contact electrode was held at ground; the microscopic Li concentration
profile and partially active transition layer were not resolved explicitly
in the electrostatic model. The calculation shown here was evaluated at the
AK01 operating bias of $700\,\mathrm{V}$ and is used to interpret the bulk
field topology rather than to fit a precise depletion voltage.

\begin{figure}[htbp]
\centering
\includegraphics[width=0.75\textwidth]{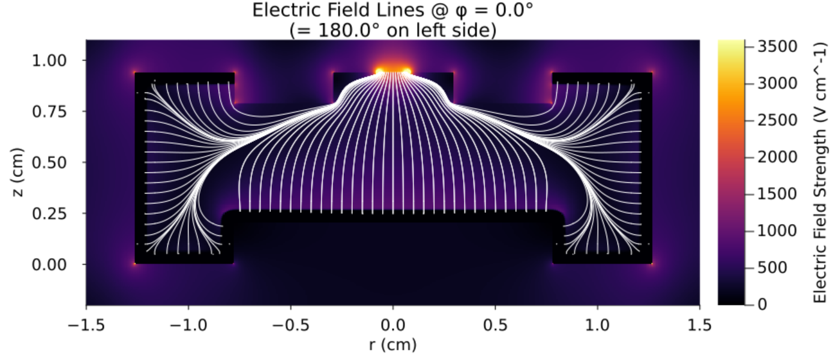}
\caption{Simulated electric-field magnitude and field-line distribution
for AK01 at an applied bias of $700\,\mathrm{V}$.}
\label{fig:electric_field}
\end{figure}

Figure~\ref{fig:electric_field} shows the simulated electric-field
magnitude together with the electric-field lines for AK01. The color
scale represents the magnitude of the electric field, while the white
curves indicate the electric-field lines within the detector geometry.

The electric field is strongly concentrated in the vicinity of the
point-contact electrode, where the field lines converge toward the
small contact. Away from the point contact, the field lines extend
through the detector bulk toward the outer contact surfaces. This
configuration is characteristic of the ICPC geometry, with a localized
high-field region near the point contact and a lower field magnitude
over much of the detector volume.

For the applied bias polarity, holes generated in the p-type bulk drift
toward the grounded point contact, while electrons drift toward the
positively biased lithium-diffused outer electrode. The field lines shown
in Fig.~\ref{fig:electric_field} therefore describe the electrostatic field
topology rather than trajectories of both carrier species toward the point
contact. Because the weighting potential of a point-contact detector is
strongly localized near the small readout electrode, hole motion near the
point contact makes a particularly important contribution to signal
formation~\cite{luke1989,comellato2021}. A systematic simulated C--V/depletion-voltage sensitivity
scan over impurity concentration and Li-transition-layer assumptions was
not part of the archived calculation and is not claimed here.

\section{Geant4 simulation}

Geant4 simulations were performed to investigate the gamma-ray response
of the fabricated ICPC detector. Since AK01 and AK02 have similar
geometrical configurations, AK02 was selected as the representative
detector for the simulation study.

\subsection{Detector geometry and simulation configuration}

The detector geometry was implemented using the measured dimensions of
AK02 together with the surrounding cryostat components used during the
experimental measurements. The archived model used a Li-diffused inactive
layer of $0.8\,\mathrm{mm}$. This value is an illustrative baseline
assumption, not a thickness measured for AK02 and not a value derived
uniquely from the nominal 280~$^\circ$C/30-min diffusion recipe. Because
the present fabrication record does not contain an independent Li
concentration-depth profile or calibrated dead-layer measurement, the
Geant4 study is not used to determine or validate the inactive-layer
thickness. A thickness-sensitivity scan was not part of the archived
simulation; at 662~keV the present spectrum is used primarily to illustrate
the detector response rather than to constrain sub-millimeter changes in
the inactive layer.

The simulations were performed using the Geant4 simulation toolkit~\cite{agostinelli2003}
with the Livermore electromagnetic physics model. A monoenergetic
$662\,\mathrm{keV}$ gamma-ray point source was positioned above the
detector and emitted gamma rays isotropically. A total of
$2\times10^{7}$ primary events were generated.

\begin{figure}[htbp]
\centering
\includegraphics[width=0.70\textwidth]{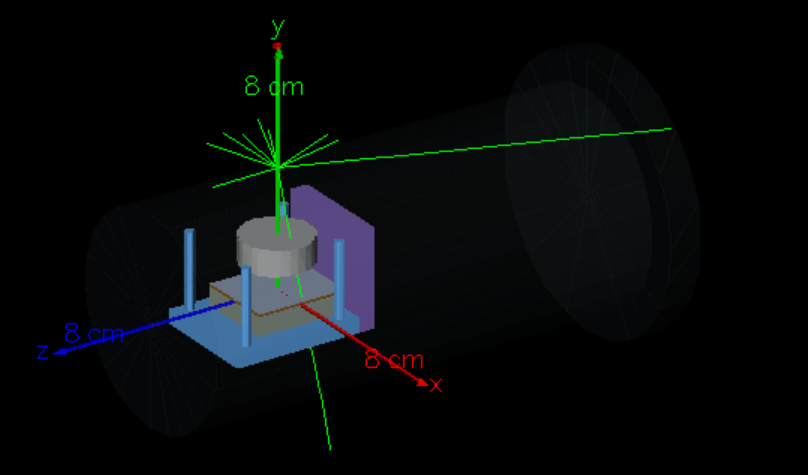}
\caption{Geant4 visualization of AK02 and the surrounding cryostat
geometry. The green lines indicate representative trajectories of
$662\,\mathrm{keV}$ gamma rays emitted isotropically from the point
source positioned above the detector.}
\label{fig:geant4_geometry}
\end{figure}

Figure~\ref{fig:geant4_geometry} shows the detector and cryostat geometry
used in the simulation together with representative gamma-ray
trajectories.

\subsection{Energy-deposition analysis and detector-response smearing}

The active germanium region was defined as the sensitive detector
volume, and the total energy deposited in each event was recorded.
The Geant4 output was analyzed using Python with the \texttt{uproot},
\texttt{awkward}, and \texttt{NumPy} libraries.

Because the ideal Monte Carlo energy-deposition spectrum does not
include the finite detector energy resolution, an energy-dependent
Gaussian broadening was applied to each simulated event. The archived
analysis used the empirical parameterization

\begin{equation}
\mathrm{FWHM}(E)=p_{0}+p_{1}E,
\end{equation}

where the parameters were determined from the measured AK02 energy
resolutions at 59.5~keV from $^{241}$Am and 661.7~keV from $^{137}$Cs.
Using the measured FWHM values of 1.04~keV at 59.5~keV and
1.62~keV at 661.7~keV gives

\[
p_{0}=0.9827~\mathrm{keV},
\qquad
p_{1}=9.63\times10^{-4}.
\]

This two-point linear relation was used as an empirical
energy-dependent resolution model for smearing the simulated AK02
$^{137}$Cs spectrum. Because the parameterization is based on only two
measured energy points, it is not intended as a globally validated HPGe
resolution model.

The corresponding Gaussian width was calculated from

\begin{equation}
\sigma(E)=\frac{\mathrm{FWHM}(E)}{2.355}.
\end{equation}

\subsection{Simulated $^{137}$Cs response}

\begin{figure}[htbp]
\centering
\includegraphics[width=0.85\textwidth]{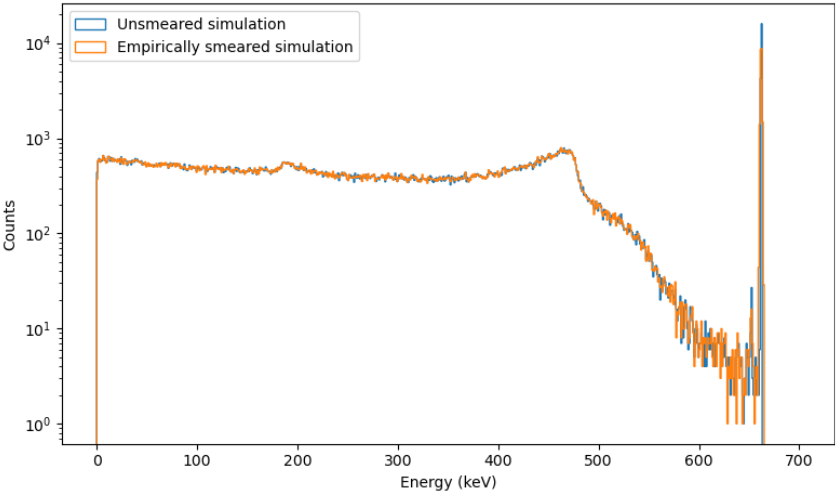}
\caption{Geant4-simulated $^{137}$Cs energy spectrum for AK02 using
an illustrative assumed $0.8\,\mathrm{mm}$ Li-diffused inactive layer. The unsmeared
energy-deposition spectrum and the energy-dependent Gaussian-smeared
spectrum are shown.}
\label{fig:geant4_spectrum}
\end{figure}

Figure~\ref{fig:geant4_spectrum} compares the unsmeared Geant4
energy-deposition spectrum with the spectrum obtained after applying the
energy-dependent Gaussian broadening. The simulated response exhibits
the characteristic $^{137}$Cs spectral features, including the Compton
continuum, the Compton edge near $477\,\mathrm{keV}$, and the
full-energy peak near $662\,\mathrm{keV}$.

A Gaussian fit to the smeared full-energy peak yielded a centroid of
\(662.01\,\mathrm{keV}\) and a standard deviation of
\(0.697\,\mathrm{keV}\), corresponding to a FWHM of
\(1.64\,\mathrm{keV}\). This is close to the measured AK02
$^{137}$Cs resolution of \(1.62\,\mathrm{keV}\) at
\(661.7\,\mathrm{keV}\). The agreement is expected because the measured
AK02 resolutions at \(59.5\,\mathrm{keV}\) and \(661.7\,\mathrm{keV}\)
were used to construct the empirical energy-dependent smearing relation.
Therefore, the comparison serves as a consistency check of the smearing
implementation rather than an independent prediction of the detector
resolution.

Figure~\ref{fig:geant4_spectrum} shows the ideal and Gaussian-smeared
Monte Carlo spectra. It is not intended as an absolute-efficiency or
full spectral-shape validation against the experimental data. A direct
measured-versus-simulated comparison of the continuum shape,
peak-to-Compton ratio, and low-energy efficiency would require a common
normalization and a calibrated source geometry and is beyond the
quantitative scope of the present simulation.

\section{Discussion and conclusion}

The fabrication and characterization results demonstrate a practical
hybrid-contact processing route for compact p-type ICPC HPGe detectors
made from USD-grown crystals. The novelty of the present study is not
lithium diffusion itself, which is established HPGe technology, but its
controlled implementation together with an a-Ge/Al point contact in two
independently fabricated ICPC prototypes. Both AK01 and AK02 exhibited
stable cryogenic operation with picoampere-level leakage currents and
well-defined operational depletion behavior. AK02 shows a steeper
high-bias leakage-current slope than AK01, so the two devices should be
viewed as successful prototypes rather than as evidence that all contact
or surface-current contributions are identical.

Gamma-ray spectroscopy demonstrated good prototype performance. At the
$^{137}$Cs full-energy peak near $662\,\mathrm{keV}$, the archived FWHM
values are $1.71\,\mathrm{keV}$ and $1.62\,\mathrm{keV}$ for AK01 and
AK02, respectively; without formal fit uncertainties, these values are
treated as comparable rather than as evidence that AK02 is intrinsically
superior. AK02 was additionally characterized with a $^{241}$Am source
and yielded a FWHM of $1.04\,\mathrm{keV}$ at $59.5\,\mathrm{keV}$.
Quadrature subtraction of the pulser contribution gives approximate
effective non-pulser contributions, but because the pulser and gamma
widths are close and fit covariances are unavailable, those residuals are
not used as precision detector-physics observables.

The electric-field simulation of AK01 shows the characteristic ICPC
field topology and provides a qualitative check that the measured geometry
supports charge transport at the operating bias. The Geant4 simulation
was performed for AK02 using its measured detector geometry and reproduces
the expected qualitative $^{137}$Cs energy-deposition features, but its
0.8-mm inactive layer is an illustrative assumption and its resolution
smearing is experimentally constrained. The simulation is therefore used
as a response-model demonstration rather than as an independent
determination of either the detector resolution or Li inactive-layer
thickness.

Overall, the results establish the fabrication and stable operation of
two compact USD-grown ICPC HPGe prototypes using Li-diffused outer
contacts and a-Ge/Al point contacts. The existing $^{241}$Am measurement
already provides a low-energy spectroscopic probe at 59.5~keV, but it was
not performed here as an absolute-efficiency measurement with the source
activity and geometry calibrated for dead-layer extraction. A future
calibrated efficiency analysis of this low-energy response, supplemented
by additional source energies or geometries and a Li-thickness sensitivity
scan in Geant4, would provide a substantially stronger quantitative
constraint on the effective inactive/transition layer than the present
662-keV response model.

\section*{Acknowledgments}

This work was supported in part by the U.S. National Science Foundation
under Grants No. OISE-1743790, OIA-2437416, and PHYS-2310027, and by
the U.S. Department of Energy under Grants No. DE-SC0024519 and
DE-SC0004768. This research was also supported by a research center
funded by the State of South Dakota. Additional support was provided by
the U.S. Air Force Research Laboratory under Award No.
FA9550-23-1-0495. We acknowledge Lawrence Berkeley National Laboratory
for providing the cryostat used for detector characterization.

\end{document}